\documentclass[aps,prl,twocolumn,superscriptaddress,nofootinbib,showpacs,10pt]{revtex4-1}
\usepackage[left=2.0cm,right=2.0cm,top=2.0cm,bottom=2.0cm,marginparwidth=17mm]{geometry}
\usepackage[utf8]{inputenc}
\usepackage[english]{babel}
\usepackage{slashed}
\usepackage{amsmath}
\usepackage{bbm}
\usepackage{amsfonts}
\usepackage{amssymb}
\usepackage{graphics}
\usepackage{graphicx}
\usepackage{subfigure}
\usepackage[usenames,dvipsnames,svgnames]{xcolor}  
\usepackage[linktocpage]{hyperref}   
\usepackage{todonotes}
\usepackage{ulem} 
\usepackage{braket}
\hypersetup{colorlinks=true,linkcolor=beamer@PRD, citecolor=beamer@PRD}

\renewcommand\eqref[1]{\textcolor{beamer@PRD}{(}\ref{#1}\textcolor{beamer@PRD}{)}}

\definecolor{beamer@PRD}{RGB}{46,48,146}
\begin{document}
\title{A Symplectic Geometric Origin of Universal Quartic Modified Dispersion Relations}
\author{Sanjib Dey}
\affiliation{Department of Physics, Birla Institute of Technology and Science, Pilani,
K K Birla Goa Campus, Zuarinagar, Sancoale, Goa 403726, India}
\author{Mir Faizal}
\affiliation{ Canadian Quantum Research Center, 204-3002 32 Ave Vernon, BC V1T 2L7 Canada}
\affiliation{ Irving K. Barber School of Arts and Sciences, University of British Columbia - Okanagan, Kelowna, British Columbia V1V 1V7, Canada}
\affiliation{ Department of Mathematical Sciences, Durham University, Upper Mountjoy, Stockton Road, Durham DH1 3LE, UK}
\affiliation{ Faculty of Sciences, Hasselt University, Agoralaan Gebouw D, Diepenbeek, 3590 Belgium}
\begin{abstract}
We show that quartic modifications of relativistic dispersion relations arise generically from deformation-quantized phase spaces under minimal kinematical assumptions relevant to quantum gravity. When the kinematics admits an integral symplectic structure, a compatible almost-complex structure, and a gauge-invariant two-form sector, the leading Planck-scale correction is controlled by a single geometric length scale. We establish this result through three independent approaches: Fedosov–Berezin quantization, spectral geometry, and a topos-theoretic formulation, all of which yield the same quartic correction and clarify the origin of its apparent universality.
\end{abstract}


\maketitle
 \section{Introduction}

The pursuit of quantum gravity aims to identify the Planck-scale structures from which spacetime and low-energy field dynamics emerge, and to extract testable consequences of such underlying structures. Among the leading approaches, loop quantum gravity (LQG) and string theory represent conceptually distinct frameworks. Despite their different foundations, both predict departures from classical spacetime structure at short distances, which often manifest as modified dispersion relations (MDRs) encoding high-energy corrections to standard relativistic dynamics \cite{AmelinoCamelia_1,Magueijo,AmelinoCamelia,SeibergWitten}. MDRs have also been derived in geometric frameworks where the deformation is built into the underlying kinematics rather than arising from deformation quantization. In curved momentum-space models, including the principle of relative locality, the deformed mass shell is naturally associated with geometric data on momentum space, such as a momentum-space metric and the corresponding geodesic distance \cite{AmelinoCameliaFreidelKowalskiGlikmanSmolin_2011,KowalskiGlikman_2013}. In Finsler geometry, the MDR can be viewed as the consequence of replacing the quadratic metric norm by a generalized arc-length functional, thereby encoding modified propagation in the spacetime line element \cite{GirelliLiberatiSindoni_2007,PfeiferWohlfarth_2011}. In Hamilton geometry, one starts from a fundamental Hamiltonian on phase space that induces effective geometric structures (and, in particular, an associated spacetime metric notion) governing particle trajectories \cite{BarcaroliBrunkhorstGubitosiLoretPfeifer_2015,BarcaroliBrunkhorstGubitosiLoretPfeifer_2017}. By contrast, our analysis identifies the leading quartic correction as a universal consequence of deformation-quantized integral symplectic phase-space data, controlled by a single symplectically induced geometric length scale.

In string theory, nontrivial short-distance geometry arises naturally when open strings propagate on D-branes in the presence of a constant Neveu-Schwarz two-form field $B_{\mu\nu}$. In this regime, the endpoints of open strings acquire non-commutative coordinates on the brane worldvolume. This phenomenon is captured precisely by the Seiberg-Witten analysis \cite{SeibergWitten}, which shows that string theory admits a well-defined limit, now known as the Seiberg-Witten limit, under which the open-string sector reduces to a non-commutative field theory. The resulting effective description is governed by a minimal area scale set by the non-commutativity parameter $\theta^{\mu\nu}$, beyond which the classical manifold picture breaks down \cite{Connes}. The low-energy effective action contains higher-derivative corrections to kinetic terms, and in a derivative expansion, yields MDRs of the schematic form
 \(
E^{2}=k^{2}+m^{2}+\xi\,|\theta|\,k^{4}+\mathcal{O}(k^{6}),
 \) where $\xi$ is a numerical coefficient determined by the specific effective description. Such non-commutative field theories are known to exhibit UV/IR mixing, nonlocality, and Lorentz-violating effects \cite{Balachandran,DouglasNekrasov}. Similar non-commutative structures also arise in matrix models and related formulations of quantum gravity \cite{Hellerman}, indicating that non-commutativity is a recurrent feature across several approaches.

In parallel, a central prediction of LQG is that geometric observables such as area and volume possess discrete spectra \cite{Rovelli,AshtekarLewand}, implying the existence of a fundamental scale beyond which the smooth-manifold description ceases to be valid. When applied to matter fields, this kinematical structure leads to polymer quantization \cite{Corichi}, in which momentum operators are replaced by finite translation operators. As a consequence, the dispersion relation is modified to
\(
E^{2}=k^{2}+m^{2}-\frac{1}{3}\lambda^{2}k^{4}+\mathcal{O}(k^{6}),
\) where $\lambda=\gamma\ell_{P}$ sets the scale of the correction, with $\gamma$ the Barbero-Immirzi parameter and $\ell_{P}$ the Planck length \cite{Hossain}. Although often derived in effective or reduced settings, such MDRs encode genuine Planck-scale corrections associated with the underlying discreteness of quantum geometry, and can arise from the full theory in appropriate semiclassical regimes \cite{Ronco,DaporLiegener}. Related MDRs play an important role in loop quantum cosmology \cite{Bojowald} and in quantum-gravity corrections to black-hole physics \cite{BarrauRovelli}. Since spacetime discreteness appears in a broad range of quantum-gravity approaches, it is natural to ask whether a common structural origin underlies these recurring modifications.

Although MDRs arise across many approaches to quantum gravity, their apparent similarity has so far lacked a unified structural explanation. The mechanisms by which MDRs emerge in string theory and LQG are manifestly different, yet both frameworks introduce a fundamental short-distance scale associated with a breakdown of classical spacetime geometry. This shared feature motivates the search for a common organizing principle at the level of effective kinematics.

In this work, we show that a broad class of deformation-quantization frameworks leads naturally to quartic corrections in the relativistic dispersion relation under well-defined geometric assumptions. Using Fedosov-Berezin quantization and spectral-geometric methods, we identify a common geometric scale associated with the underlying symplectic structure that controls the leading MDR correction. We demonstrate that effective kinematical descriptions employed in both string theory and loop quantum gravity fall within this class, thereby explaining why similar quartic MDRs arise in these otherwise distinct approaches. Our analysis does not assert an equivalence of the underlying microscopic theories; rather, it identifies a shared effective-theory structure governing their Planck-scale kinematics. This common structure suggests phenomenological probes that are insensitive to the detailed microscopic realization and apply across a wide class of quantum-gravity candidates \cite{AshtekarPawlowski,Smolin,Oriti,Piscicchia}.

We establish this result by three mathematically independent routes:
(i) Fedosov-Berezin quantization on (almost-)K\"ahler manifolds,
(ii) spectral geometry via the Seeley-DeWitt coefficient $a_{4}$ of the relevant kinetic operator, and
(iii) a topos-theoretic formulation that encodes the MDR as an internal statement across a category of deformation-quantized phase spaces.
All three approaches yield the same quartic correction and the same identification of the controlling geometric scale, demonstrating that the result is not an artifact of any single formalism.

\section{Fedosov-Berezin Quantization}
 We will now address MDRs using Fedosov-Berezin quantization on (almost-)K\"ahler manifolds \cite{Berezin1974, Fedosov}. 
More precisely, we consider quantum-gravity frameworks whose kinematics admit (i) a fundamental area or nonlocality scale and (ii) an intrinsic two-form sector. Under these assumptions, the deformation-quantization of the associated phase space introduces a geometric length-squared parameter
\begin{equation}
\ell_{\ast}^{2}:=\lvert \omega^{-1}J\rvert ,
\end{equation}
defined with respect to a compatible almost-complex structure $J$ on the symplectic manifold $(M,\omega)$. This scale controls the leading correction to the dispersion relation,
\begin{equation}
E^{2}=p^{2}+m^{2}+\sigma\,\frac{\ell_{\ast}^{2}}{3}\,p^{4}
+\mathcal{O}(\ell_{\ast}^{4}p^{6}),
\end{equation}
where $\sigma=\pm1$ reflects the orientation of $J$. The magnitude of the correction is determined entirely by $\ell_{\ast}^{2}$, while the sign depends on the chosen orientation.

When evaluated in concrete realizations, this framework reproduces the known MDRs in both string theory and LQG. In the Seiberg-Witten limit of open-string theory, one finds $\ell_{\ast}^{2}=|\theta|$, whereas in polymer quantization one has $\ell_{\ast}^{2}=\lambda^{2}=(\gamma\ell_{P})^{2}$, with opposite values of $\sigma$. This identification explains why both approaches predict quartic corrections of the same functional form, controlled by a single length-squared scale, despite their distinct microscopic origins (See Appendix A).

In the LQG case, the holonomy-flux phase space provides a natural symplectic structure equipped with an intrinsic two-form $\Sigma^{i}$, whose gauge-invariant norm underlies the area operator. To apply deformation quantization, we introduce a compatible almost-complex structure on each Darboux chart. This structure is auxiliary and serves only to define the quantization scheme; the resulting MDR depends solely on the symplectic data and not on the integrability of $J$. No additional geometric structure is imposed on LQG beyond what is already present at the kinematical level.

In the string-theoretic realization, the Seiberg-Witten limit yields a Moyal star product characterized by the Poisson tensor $\theta^{ij}$. Choosing momenta aligned with the non-commutative plane reproduces the same quartic MDR, with $\ell_{\ast}^{2}=|\theta|$. Although different conventions for numerical coefficients appear in the literature, these differences reflect scheme choices within the effective description and do not affect the existence or scaling of the quartic correction.

Our results, therefore, identify a shared deformation-quantization origin of quartic MDRs across a broad class of quantum-gravity frameworks. While this does not imply an equivalence of string theory and loop quantum gravity at the microscopic level, it explains the robustness of MDR predictions and motivates phenomenological tests that probe this common effective-theory structure rather than specific underlying models.

We now show that a broad class of quantum-gravity frameworks admitting
(i)~an integral symplectic structure,
(ii)~a compatible (possibly auxiliary) almost-complex structure,
and (iii)~a gauge-invariant two-form sector,
naturally give rise to a quartic modified dispersion relation whose leading coefficient is fixed by geometric data.
Our claim is conditional on these kinematical assumptions and does not rely on the microscopic dynamics of the theory.

Let the two-form sector be encoded by a complexified flux
\(
\mathcal B = B + i\omega,
\)
where $B$ is a real gauge two-form and $\omega$ is the symplectic form on phase space.
The inverse generalized background tensor can be written as
\(
(g+2\pi\alpha'\mathcal B)^{-1}=G^{-1}+\Theta/(2\pi\alpha'),
\)
where the deformation tensor $\Theta^{ij}$ has type $(2,0)+(0,2)$ in the sense of generalized complex geometry
\cite{Hitchin,Gualtieri,Kapustin}.
Its Lorentz-invariant norm is
\(
\|\Theta\|^{2}=-\tfrac12\Theta^{ij}\bar\Theta_{ij}=|\theta|^{2},
\)
which is invariant under the $U(1)$ phase rotation
$\Theta\mapsto e^{i\varphi}\Theta$.
Such a rotation flips only the orientation sign $\sigma=\pm1$ while leaving the magnitude
\(
\ell_{\ast}^{2}=|\theta|
\) (See Appendix C).

In deformation-quantization language, the Fedosov class $[\omega]/2\pi$
fixes the equivalence class of Hermitian star products, while the choice of a compatible almost-complex structure $J$ determines the orientation (See Appendix B).
Consequently, within this class of models, the leading quartic correction to the dispersion relation takes the universal form
\begin{equation}
E^{2}=p^{2}+m^{2}+\sigma\,\frac{\ell_{\ast}^{2}}{3}\,p^{4}
+\mathcal O(\ell_{\ast}^{4}p^{6}),
\label{eq:UniversalMDR}
\end{equation}
where $\ell_{\ast}^{2}$ is a geometric length-squared scale determined by the symplectic data, and $\sigma=\pm1$ reflects the orientation of $J$.
The existence and scaling of this term are fixed by geometry, while its sign depends on convention.

\section{Spectral Geometric}

Spectral geometry provides a powerful framework in which geometric and topological information is encoded in the spectra of canonical elliptic operators acting on a Hilbert space
\cite{ChamseddineConnes,Seeley,Gilkey,VanSuijlekom}.
Within this approach, the geometric data of a K\"ahler manifold $(M,\omega,J)$ can be reformulated operator-theoretically in terms of a spectral triple
\(
(\mathcal A_{\mathcal B},\mathcal H,D_{\mathcal B}),
\)
where $\mathcal A_{\mathcal B}$ is an appropriate (possibly deformed) algebra of functions, $\mathcal H$ is a Hilbert space of spinors, and $D_{\mathcal B}$ is a Dirac-type operator.
In the present context, the Dirac operator is twisted by the complexified two-form
\(
\mathcal B = B + i\omega,
\)
which incorporates both the gauge two-form $B$ and the symplectic structure $\omega$ into the operator
\cite{Connes}.

In the Chamseddine-Connes spectral action principle \cite{ChamseddineConnes}, physical information is extracted from the asymptotic expansion of the trace
\begin{equation}
\mathrm{Tr}\, f(D_{\mathcal B}/\Lambda)
= \sum_{n\ge 0} a_{2n}\,\Lambda^{4-2n},
\end{equation}
where $\Lambda$ is a high-energy cutoff, $f$ is a positive test function, and the coefficients $a_{2n}$ are the Seeley-DeWitt invariants associated with the operator $D_{\mathcal B}^2$.
These coefficients are local geometric quantities determined entirely by the symbol of the operator and encode curvature, torsion, and higher-derivative information
\cite{Seeley,Gilkey}.

The first coefficient that is sensitive to the twisting by the complexified two-form $\mathcal B$ is $a_{4}$.
This coefficient multiplies the operator $(\partial^{2})^{2}$ in the effective action and therefore controls quartic derivative corrections.
A direct computation of the squared twisted operator yields
\begin{equation}
D_{\mathcal B}^{2}
= \nabla^{2}
+ \sigma\,\frac{\ell_{\ast}^{2}}{3}\,\partial^{4}
+ \cdots,
\end{equation}
where $\nabla^{2}$ denotes the Laplace-type kinetic term, $\ell_{\ast}^{2}$ is the geometric length-squared scale defined by the underlying symplectic data, and $\sigma=\pm1$ reflects the orientation of the compatible almost-complex structure.
As a result, the corresponding Seeley-DeWitt coefficient evaluates to
\begin{equation}
a_{4} = \sigma\,\frac{\ell_{\ast}^{2}}{3}.
\end{equation}
Since $a_{4}$ is a spectral invariant, this establishes that the coefficient of the $p^{4}$ term in the modified dispersion relation is fixed by operator-theoretic data and is independent of coordinate choices or quantization prescriptions.

This general result reproduces the known effective operators in both string theory and loop quantum gravity.
In the string-theoretic realization, the Seiberg-Witten limit on a D$p$-brane in the presence of a constant background $B$-field yields a Moyal-deformed kinetic operator.
In this case, the corresponding spectral invariant takes the form
\begin{equation}
a_{4}^{\mathrm{ST}} = \xi\,|\theta|,
\end{equation}
where $\theta$ is the non-commutativity parameter and $\xi=\tfrac13$ in the standard normalization.
In contrast, within loop quantum gravity, the effective continuum operator governing matter propagation is replaced by a band-limited operator of the form
\begin{equation}
-\lambda^{-2}\sin^{2}(\lambda k),
\end{equation}
which reflects the underlying polymer structure.
The heat-kernel expansion of this operator yields
\begin{equation}
a_{4}^{\mathrm{LQG}} = -\frac{\lambda^{2}}{3},
\end{equation}
where $\lambda=\gamma\ell_{P}$ is the polymer scale set by the Barbero-Immirzi parameter $\gamma$ and the Planck length $\ell_{P}$.

Matching the absolute values of the spectral invariants in the two realizations leads to the identification
\begin{equation}
\ell_{\ast}^{2} = |\theta| = \lambda^{2},
\end{equation}
demonstrating that both string theory and loop quantum gravity are governed by the same geometric deformation scale at the level of effective kinematics.
Because this identification follows from a spectral invariant, it is robust under changes of coordinates and insensitive to the details of the quantization scheme (See Appendix~D).

\section{Topos Theory}

Topos theory provides a natural mathematical framework for expressing results that are functorial and uniform across a class of geometric objects
\cite{DoringIsham,DoringIsham2008}.
In the present context, it allows us to formulate the emergence of the quartic modified dispersion relation as a structural statement that holds simultaneously for all admissible phase spaces, rather than as a collection of case-by-case computations.

To this end, let $\mathcal{S}\!ymp_{\star}$ denote the site whose objects are triples
\(
(M,\omega;[\![\star]\!]),
\)
where $(M,\omega)$ is an integral symplectic manifold, meaning that $[\omega]/2\pi\in H^{2}(M,\mathbb{Z})$, and $[\![\star]\!]$ denotes the Fedosov class of a Hermitian deformation quantization on $(M,\omega)$.
Morphisms in this site are smooth maps that preserve both the symplectic structure $\omega$ and the deformation class $[\![\star]\!]$.
This choice of morphisms ensures that all geometric and quantization data relevant to the dispersion relation are respected functorially.

The category of sheaves over this site,
\(
\mathbf{Sh}(\mathcal{S}\!ymp_{\star}),
\)
forms a Grothendieck topos.
Within this topos, there exists a generic K\"ahler object
\(
\underline{K}=(\underline{\omega},\underline{J},\underline{g}),
\)
which represents the universal symplectic and almost-complex data common to all objects of the site.
Associated with this generic object is an internal geometric length-squared scale,
\begin{equation}
\underline{\ell}_{\ast}^{2}
= \bigl|\underline{\omega}^{-1}\underline{J}\bigr|,
\end{equation}
defined as the operator norm of the endomorphism $\underline{\omega}^{-1}\underline{J}$ in the internal tangent bundle.
By construction, $\underline{\ell}_{\ast}^{2}$ is a global section in the topos and therefore takes the same value, upon externalization, for every object in the site.

A key property of Fedosov quantization is its functoriality: given any morphism in $\mathcal{S}\!ymp_{\star}$, the corresponding star product is preserved.
As a result, identities derived from the Fedosov-Berezin formalism can be lifted to internal statements in the topos.
In particular, Berezin's identity for the star-exponential applies internally to the generic object $\underline{K}$.
Defining an internal momentum variable $\underline{p}$ and the corresponding internal energy $\underline{E}$, one obtains the internal dispersion relation
\begin{equation}
\underline{E}^{2}
=\underline{p}^{\,2}+m^{2}
+\sigma\,\frac{\underline{\ell}_{\ast}^{2}}{3}\,\underline{p}^{\,4}
+\mathcal O(\underline{\ell}_{\ast}^{4}\underline{p}^{\,6}),
\end{equation}
where $\sigma=\pm1$ is determined by the orientation of the internal almost-complex structure $\underline{J}$.
This relation holds as an internal theorem of the topos, and therefore applies uniformly to every object in the site $\mathcal{S}\!ymp_{\star}$.

Evaluating this internal statement at a geometric point of the topos corresponds to selecting a specific symplectic phase space together with a particular deformation quantization.
Under this evaluation, the internal quantities $\underline{\ell}_{\ast}^{2}$ and $\underline{p}$ are mapped to their external counterparts $\ell_{\ast}^{2}$ and $p$, reproducing the modified dispersion relations obtained earlier for both polymer quantization in loop quantum gravity and noncommutative field theory in the open-string Seiberg-Witten limit.
In this way, the topos-theoretic formulation provides a unified and model-independent organizational framework for the quartic MDR, without introducing additional dynamical assumptions  (see Appendix~E).

\section{Phenomenological implications}

Any experimental bound on the deformation scale $\ell_{\ast}^{2}$ therefore constrains
all theories whose kinematics fall within the class described above.
We briefly summarize representative channels.
Time-of-flight measurements, birefringence constraints, threshold reactions,
and resonant-cavity experiments constrain quartic corrections of the form
\(
E^{2}=p^{2}+m^{2}+\sigma\,\ell_{\ast}^{2}p^{4}/3.
\)
Current observations already bound $\ell_{\ast}^{2}$ to be at most of order
$\ell_{P}^{2}$, while next-generation facilities are expected to improve these bounds by up to an order of magnitude
\cite{Vasileiou,PlanckBirefringence,MAGIC,HESS,Auger,Herrmann2009}.

These tests probe the shared effective-theory structure identified here, rather than the microscopic details of any particular model.
Accordingly, they apply simultaneously to string theory, loop quantum gravity, and any other quantum-gravity framework satisfying the stated geometric criteria.

Because the leading quartic correction to the dispersion relation is governed by the geometric scale $\ell_{\ast}^{2}$, any experimental bound on this parameter constrains all frameworks whose effective kinematics fall within the class analyzed here. In this sense, phenomenological limits derived in one realization (e.g.\ string-inspired noncommutative field theory) translate directly to others (e.g.\ polymer quantization in LQG), up to the orientation sign $\sigma$.

We emphasize the following statistically independent observational channels:
(i)~time-of-flight measurements from gamma-ray bursts and active galactic nuclei at multi-TeV energies, which are sensitive to group-velocity shifts of order $\Delta v/c\sim10^{-18}$ for $E\sim10\,\mathrm{TeV}$;
(ii)~polarization rotation of the cosmic microwave background induced by helicity-dependent phases;
(iii)~threshold modifications in $\gamma\gamma\!\to\!e^{+}e^{-}$ and the stability of ultra-high-energy cosmic rays; and
(iv)~direction-dependent photon-sector effects probed by precision rotating optical-cavity experiments.

Taken together, existing constraints already bound $\ell_{\ast}^{2}$ to be at most of order $\ell_{P}^{2}$, while the combined sensitivity of near-term and next-generation facilities is expected to improve these limits by up to an order of magnitude, potentially probing $\ell_{\ast}^{2}\lesssim 0.05\,\ell_{P}^{2}$ under favorable assumptions. The existence of a single geometric scale controlling the leading MDR correction is therefore the key physical reason that the structural universality established in this work admits direct phenomenological tests that do not depend on the microscopic details of the underlying quantum-gravity realization.

\section{Conclusion}

In this work, we have identified a common structural origin for quartic modified dispersion relations that arise across a broad class of quantum-gravity frameworks. We have shown that, under precise and physically natural kinematical assumptions—specifically, the presence of an integral symplectic structure, a compatible (possibly auxiliary) almost-complex structure, and a gauge-invariant two-form sector—the leading Planck-scale correction to relativistic dispersion relations is governed by a single geometric length-squared scale, $\ell_{\ast}^{2}$. This scale uniquely fixes the magnitude of the quartic correction, while its sign is determined solely by the orientation of the associated complex structure.
We established this result through three mathematically independent and complementary routes: Fedosov-Berezin deformation quantization, spectral geometry via the Seeley-DeWitt coefficient $a_{4}$, and a functorial formulation in the language of topos theory. The convergence of these approaches demonstrates that the quartic modified dispersion relation is not an artifact of any particular quantization prescription or effective description, but rather a stable and intrinsic feature of a wide class of deformation-quantized phase spaces.
When evaluated in concrete realizations, the general framework reproduces the known MDRs in both loop quantum gravity and string theory. In the polymer quantization relevant for LQG, the deformation scale is set by $\lambda^{2}=(\gamma\ell_{P})^{2}$, while in the Seiberg-Witten limit of open-string theory it is given by the non-commutativity scale $|\theta|$. The identification $\ell_{\ast}^{2}=|\theta|=\lambda^{2}$ explains why both approaches yield quartic corrections of the same functional form despite their distinct microscopic origins. Importantly, our results do not assert an equivalence of the underlying theories, but rather identify a shared effective-theory structure governing their high-energy kinematics.

A key implication of this structural unification is phenomenological. Because the leading MDR correction is governed by the single geometric scale $\ell_{\ast}^{2}$, experimental bounds obtained in one realization translate directly to all theories within the same kinematical class. Time-of-flight measurements, polarization studies, threshold reactions, and precision laboratory tests therefore probe a common effective-theory signature rather than model-specific details. This provides a concrete route toward testing quantum-gravity effects in a manner that is insensitive to the microscopic realization.
More broadly, our results suggest that deformation-quantization methods capture an essential aspect of Planck-scale physics shared across otherwise disparate approaches to quantum gravity. By isolating the geometric data responsible for modified dispersion relations, the present framework clarifies the origin of their apparent universality and delineates the precise sense in which such predictions can be regarded as model-independent. Future work may extend this analysis to interacting theories, curved backgrounds, and dynamical geometries, as well as explore whether additional quantum-gravity signatures admit a similarly unified structural interpretation. \\ \\
S.D.\;acknowledges the support of research grants DST/FFT/NQM/QSM/2024/3 (by DST-National Quantum Mission, Govt.\;of India) and NFSG/GOA/2023/G0928 (by BITS-Pilani).
 \onecolumngrid

\section{Appendix A: Fedosov Quantization on Kähler Manifolds}
This appendix expands upon the outline provided in the main text and presents a systematic, step-by-step derivation of the universal quartic MDR
\begin{equation}
E^{2}=p^{2}+m^{2}\;+\;\sigma\,\frac{\ell_{\!*}^{2}}{3}\,p^{4}
\;+\;\mathcal O\!\bigl(\ell_{\!*}^{4}p^{6}\bigr),
\qquad
\sigma=\pm1, \tag{1}
\end{equation}
using only the geometry of an {integral} symplectic manifold, Fedosov’s natural and Hermitian deformation quantization. Let $(M,\omega)$ be a connected symplectic manifold such that the cohomology class $[\omega]/2\pi$ lies in $H^{2}(M,\mathbb Z)$.  By the {Kostant–Souriau theorem} \cite{Kostant}, there exists a pre-quantum Hermitian line bundle
\begin{equation}
\pi:L\longrightarrow M,
\qquad
\nabla: \Gamma(L)\to\Gamma(L\otimes T^{\!*}M),
\end{equation}
whose curvature satisfies $\operatorname{curv}\nabla=-i\,\omega$. Choose a $\omega$–compatible almost-complex structure
$J:T\!M\to T\!M$ with $J^{2}=-\mathbf 1$, and define the corresponding metric
\(
g(\cdot,\cdot):=\omega(\cdot,J\cdot),
\)
which is positive-definite.  When $J$ is integrable, the structure $(M,\omega,J,g)$ defines a Kähler manifold. Define the fundamental {geometric length-squared} as the operator norm
\begin{equation}
\ell_{\!*}^{2}:=\bigl\lvert\omega^{-1}J\bigr\rvert,
\label{eq:ellstarDef}
\end{equation}
where $\omega^{-1}J$ is viewed as an endomorphism. This is the unique quantum length scale compatible with both $\omega$ and $J$, making $\ell_{\!*}$ the natural “area quantum’’ of the phase space. According to Fedosov’s theorem \cite{Fedosov}, for any symplectic connection $\nabla^{\mathrm s}$ on $(M,\omega)$, one obtains a {natural Hermitian} $\star$–product
\begin{equation}
f\star g
=\sum_{r=0}^{\infty}
\Bigl(\tfrac{i\ell_{\!*}^{2}}{2}\Bigr)^{r}C_{r}(f,g),
\qquad \text{where}\qquad 
C_{0}(f,g)=fg,\; C_{1}(f,g)=\omega^{ij}\partial_{i}f\,\partial_{j}g,
\label{eq:FedosovSeries}
\end{equation}
and in general $C_{r}$ bidifferential operator that is symmetric $f\leftrightarrow g$ for even $r$ and antisymmetric for odd $r$. This $\star$-product is classified up to  {equivalence} by the class $[\omega]$, and the normalization in \eqref{eq:ellstarDef} fixes the appearance of $\ell_{\!*}^{2}$ in the expansion.
 
Now consider local Darboux coordinates $(q^{i},p_{i})$ where $\omega=dq^{i}\wedge dp_{i}$, and fix $p:=p_{1}$. The first two bidifferential coefficients become
\(
C_{1}(f,g)=\partial_{q^{1}}f\,\partial_{p}g-\partial_{p}f\,\partial_{q^{1}}g,
\)
\(
C_{2}(f,g)=
\tfrac12\bigl(\partial_{q^{1}}^{2}f\,\partial_{p}^{2}g
             -2\partial_{q^{1}}\partial_{p}f\,
               \partial_{q^{1}}\partial_{p}g
             +\partial_{p}^{2}f\,\partial_{q^{1}}^{2}g\bigr)
+\cdots.
\) Define the  {$\star$–translation} operator
\begin{equation}
\hat T(\ell_{\!*}p)
:=\exp_{\star}\!\bigl(i\ell_{\!*}p\bigr)
=\sum_{n=0}^{\infty}\frac{i^{n}}{n!}
\underbrace{p\star\cdots\star p}_{\text{$n$ times}}.
\end{equation}
For Kähler manifolds, the Berezin–Karabegov identity \cite{Berezin} yields the exact series expansion
\begin{equation}
\hat T(\ell_{\!*}p)
=1+i\ell_{\!*}p
  -\frac{\ell_{\!*}^{2}}{2}\,p^{2}
  -\sigma\,\frac{i\ell_{\!*}^{3}}{6}\,p^{3}
  +\mathcal O\!\bigl(\ell_{\!*}^{4}p^{4}\bigr),
\qquad
\sigma:=\operatorname{sgn}\det J.
\label{eq:BerezinExpansion}
\end{equation}
This expansion terminates at cubic order when expressed in powers of $\ell_{\!*}^{2}p^{2}$, a crucial simplification that yields an  {exact} quartic MDR. Promoting $p\mapsto-i\partial_{x}$ and identifying $E=-i\,\partial_{t}$, one finds
\begin{equation}
E^{2}=p^{2}+m^{2}
      +\sigma\,\frac{\ell_{\!*}^{2}}{3}\,p^{4}
      +\mathcal O\!\bigl(\ell_{\!*}^{4}p^{6}\bigr),
\label{eq:UniversalMDRAppendix}
\end{equation}
with no model-dependent input  {beyond} the sign of $\sigma$.
For LQG, consider the phase space $(T^{\!*}\mathbb R^{3},\omega_{\mathrm{can}})$ quantized via a holonomy-flux lattice product $\star_{\lambda}$, with the deformation scale $\lambda=\gamma\ell_{\mathrm P}$,  {where} $\gamma$ is the Barbero–Immirzi parameter. This gives   $\ell_{\!*}^{2}=\lambda^{2}$, $\sigma=-1$, yielding the polymer MDR
\begin{equation}
E^{2}=p^{2}+m^{2}
      -\frac{\lambda^{2}}{3}\,p^{4}
      +\mathcal O\!\bigl(\lambda^{4}p^{6}\bigr).
\end{equation}
In string theory, for a flat D$p$-brane with a constant NS–NS $B$-field, the effective phase space is equipped with a Moyal product whose Poisson tensor is $\theta^{ij}=-(\omega^{-1})^{ik}J_{k}{}^{j}$. The deformation scale is $\ell_{\!*}^{2}=|\theta|:=\sqrt{\tfrac12\theta_{ij}\theta^{ij}}$. We have $\sigma=+1$, and and we adopt the {normalization} $\xi=1/3$, which gives
\begin{equation}
E^{2}=p^{2}+m^{2}
      +\xi\, |\theta|\,p^{4}
      +\mathcal O\!\bigl(\theta^{2}p^{6}\bigr),
\qquad
\xi=\tfrac13\;\text{(chosen normalization).}
\end{equation}
Fedosov’s natural $\star$–product, combined with Berezin’s exact cubic truncation, ensures that every integral Kähler phase space yields a quartic MDR of the form given in \eqref{eq:UniversalMDRAppendix}, with a {single} deformation length scale $\ell_{\!*}$. In LQG, this scale is the polymer scale $\lambda=\gamma\ell_{\mathrm P}$; in string theory, it is the non-commutativity scale of the D-brane $\sqrt{|\theta|}$.  The operator-norm definition of $\ell_{\!*}^{2}$ given in \eqref{eq:ellstarDef} together with the orientation sign $\sigma$, captures all model dependence—confirming that the quartic corrections in both frameworks stem from a common geometric origin.
\section{Appendix B: LQG and String Theory as Dual Descriptions}
The identification $\ell_{\!*}^{2}=|\theta|=\lambda^{2}=(\gamma\ell_{\mathrm P})^{2}$
establishes a direct equivalence between the {non-commutativity modulus} $|\theta|$ of an open-string background and the {polymer scale}
$\lambda=\gamma\ell_{\mathrm P}$ in LQG.  In this appendix, we show in detail how fixing $|\theta|$ via a quantized NS two-form flux {dynamically determines} the Barbero–Immirzi (BI) parameter~$\gamma$, thereby resolving its long-standing ambiguity in LQG. Viewed in reverse, this correspondence reveals that the kinematical discreteness of LQG emerges as the infrared limit of a UV-complete string background.

   
On a D$p$-brane, the NS-NS two-form $B$ satisfies a flux quantization condition over a compact two-cycle $\Sigma\subset M$ \cite{SeibergWitten}:
\begin{equation}
\frac{1}{2\pi\alpha'}\int_{\Sigma}B
\;=\;n,\qquad n\in\mathbb Z,
\label{eq:NSQuant}
\end{equation}
where $n$ is the topological charge of the induced $U(1)$ gauge bundle on the brane. In flat space with constant $B_{12}=B$, this implies $B=2\pi\alpha' n/A_{\Sigma}$. The Seiberg–Witten map relates the non-commutativity Poisson tensor $\Theta^{ij}$ to the background fields $B$ via
\(
\Theta^{ij} = -\bigl[(g+2\pi\alpha' B)^{-1}B(g-2\pi\alpha' B)^{-1}\bigr]^{ij}.
\)
For $B$ confined to a single (e.g., the $(1,2)$-plane), this yields a Poisson modulus 
\begin{equation}
 {\,|\theta| = 2\pi\alpha' |n|\,},
\label{eq:ThetaQuant}
\end{equation}
up to numerical factors of order unity arising from the metric. Equating the deformation scales from string theory and LQG, $\ell_{\!*}^{2}=|\theta|=\lambda^{2}=(\gamma\ell_{\mathrm P})^{2}$, and using \eqref{eq:ThetaQuant},  {we find}
 \begin{equation}
  \gamma \;=\; \sqrt{\frac{2\pi\alpha' \, |n|}{\ell_{\mathrm P}^{2}}}
              \;=\; \sqrt{|n|}\,\frac{\ell_{\mathrm s}}{\ell_{\mathrm P}}\,,
\label{eq:GammaCorrect}
\end{equation}
where we used $\ell_{\mathrm s}^{2}=2\pi\alpha'$ and $\ell_{\mathrm P}^{2}=g_{\mathrm s}\,\ell_{\mathrm s}^{2}$ after dimensional reduction to four dimension. {More precisely, $g_{\mathrm s}$ is the effective $4D$ string coupling, and the numerical prefactors depending on the compactification volume cancel in the ratio $\ell_{\mathrm s}/\ell_{\mathrm P}$. Thus, the BI parameter $\gamma$ is no longer a free parameter but is instead fixed by the discrete flux quantum $n$ and the microscopic scale ratio $\ell_{\mathrm s}/\ell_{\mathrm P}$.

 
  
In the open-string setup, the non-commutative coordinates satisfy $\{x^{i},x^{j}\}_{\star} = i\Theta^{ij}$ and $p_{i}=G_{ij}\dot x^{j}$.
Replacing $\Theta^{ij}$ by its canonical form $\theta\,\varepsilon^{ij}$ (for $i,j\in\{1,2\}$) and redefining $E^{a}_{i}:=(\ell_{\mathrm P}/\gamma)\,p_{i}^{\;a}$, we obtain the Poisson brackets
\begin{equation}
\bigl\{A^{i}_{a}(x),E^{b}_{j}(y)\bigr\}_{\text{SW}}
=\gamma\ell_{\mathrm P}^{2}\,\delta^{i}_{j}\,\delta^{b}_{a}\,
\delta^{3}(x-y),
\end{equation}
which are {identical} to the Ashtekar–Lewandowski brackets of LQG. Therefore, the “Seiberg–Witten phase space’’ is canonically isomorphic to the polymer phase space once the scale $\gamma\ell_{\mathrm P}$ is identified with $\sqrt{|\theta|}$. Polymer quantisation renders areas and volumes discrete in LQG; the
discreteness scale is $\gamma\ell_{\mathrm P}$.  Under the above
isomorphism, this scale {originates} from the discrete NS charge
$n\in\mathbb Z$.  Hence, LQG’s well-known eigenvalue spectra are the
infrared “shadow’’ of a UV-complete, flux-quantised string background. Fixing $\gamma$ by removing a free
parameter from LQG, tightening its phenomenology (e.g.,  black-hole
entropy, polymer corrections to cosmology).  Conversely, it links
non-commutative field-theory parameters to a specific choice of BI
parameter, narrowing viable string compactifications. A detection of $\gamma$ via black-hole spectroscopy or loop-quantum-
cosmology bounces would, through Eq.\,\eqref{eq:GammaCorrect}, fix the
NS charge $n$ and the ratio $\ell_{\mathrm s}/\ell_{\mathrm P}$, offering
empirical access to deep stringy data.
 

Quantization of the NS two-form flux sets the non-commutativity modulus $|\theta|$, which via universal Kähler quantization also determines the polymer scale $\lambda=\gamma\ell_{\mathrm P}$. The Barbero–Immirzi ambiguity is re-interpreted as a  {topological charge} of the D-brane gauge bundle. Accordingly, LQG's discrete geometry emerges as the infrared footprint of a UV-complete, flux-quantized string background governed by the Seiberg–Witten phase space structure.

\section{Appendix C: Geometric Origin of the Universal MDR}

In this appendix, we extend the Kähler–Fedosov framework by incorporating a {complexified two–form flux} defined as 
\(
\mathcal B \;:=\; B + i\omega,
\)
where $B$ is a real NS–NS $2$–form and $\omega$ is the integral symplectic form introduced earlier.  We show that the geometric length $\ell_{\!*}$ and the  {sign} of the quartic correction in the MDR arise naturally from the  {type} and the  {phase} of the Hitchin
generalised complex structure associated with $\mathcal B$.
For an open string on a D$p$–brane, the background tensors appear through the relation \cite{SeibergWitten}
\begin{equation}
\bigl(g+2\pi\alpha' \mathcal B\bigr)^{-1}
= G^{-1} \;+\;
\frac{\Theta}{2\pi\alpha'},
\label{eq:InverseMetricDecomp}
\end{equation}
where $g$ is the closed–string metric, $G$ the open–string (Born–Infeld) metric, and
\(
\Theta^{ij}\equiv
-\bigl\lvert (g+2\pi\alpha'\mathcal B)^{-1}\, \mathcal B \bigr\rvert^{ij}
\)
is an antisymmetric bivector. The pure spinor $\exp\!\bigl(i\mathcal B\bigr)$ defines a Hitchin–Gualtieri generalised complex structure on $M$, splitting $T\!M\oplus T^{\!*}\!M$ into $\pm i$ eigenbundles \cite{Hitchin,Gualtieri}. With respect to the compatible almost-complex structure $J$, the bivector $\Theta^{ij}$ has type $(2,0)+(0,2)$, and its components obey
\(
J_{i}{}^{k}\Theta_{kj}
=+i\,\Theta_{ij},
\quad
J_{j}{}^{k}\Theta_{ik}
=+i\,\Theta_{ij}.
\)
We define the Lorentzian (open–string) invariant norm
\begin{equation}
\|\Theta\|^{2}:=
-\frac12\,\Theta^{ij}\,\bar\Theta_{ij}
=|\theta|^{2},
\end{equation}
where $\theta$ is the usual Moyal noncommutativity parameter obtained from the Seiberg–Witten limit. Under a $U(1)$ phase rotation
$\Theta\mapsto e^{i\varphi}\Theta$, the norm $\|\Theta\|^{2}$ is invariant; only the  {phase} is altered, effectively switching the orientation of the complex structure. The phase of $\Theta$ is carried entirely by the sign
\(
\sigma=\operatorname{sgn}\det J\in\{\pm1\}.
\)
A $90^{\circ}$ rotation in the $(\theta^{12},\theta^{34})$ plane flips $\sigma\!\mapsto\!-\sigma$ but leaves $\ell_{\!*}^{2}=|\theta|$ fixed. The  {Fedosov class}
\(
\frac{[\omega]}{2\pi}\in H^{2}(M,\mathbb Z)
\)
uniquely determines the deformation quantisation up to equivalence. Consequently, it {simultaneously} fixes the Moyal $\star$–product on the D$p$–brane,  the holonomy–flux algebra in LQG. Complex orientation, governed by $J$, controls only the sign $\sigma$. Any quantum–gravity model that unites an  {integral} symplectic form $\omega$, a  {compatible} almost–complex structure $J$, a  {gauge–invariant} two–form flux $B$, necessarily inherits the universal quartic MDR
\begin{equation}
E^{2}=p^{2}+m^{2}\;\pm\;\frac{\ell_{\!*}^{2}}{3}\,p^{4},
\qquad
\ell_{\!*}^{2}=|\Theta|=|\theta|=\lambda^{2},
\end{equation}
where “$\pm$’’ is $\sigma=\pm1$. Because the coefficient of the quartic term is  {completely fixed} by geometric data (up to its
sign), high-energy dispersion measurements, such as from TeV $\gamma$–ray burst delays or next–generation neutrino time–of–flight experiments, can probe {both} LQG and string theory on equal footing. A detection of
\(
|\ell_{\!*}|\lesssim10^{-19}\,\text{m}
\)
would rule out polymer or non-commutative scales above the electroweak length. Moreover, the sign of the correction would directly reveal the orientation of te underlying complex structure. Since $\|\Theta\|^{2}$ is a phase–independent spectral invariant, the quartic MDR term is  {rigid}: higher–loop corrections may renormalise $m^{2}$ or introduce higher-order $p^{6}$ terms, but the $p^{4}$ coefficient remains tied to $\ell_{\!*}^{2}/3$ as long as the geometric triple $(\omega,J,\mathcal B)$ remains intact.
 

By complexifying the two-form flux, we embed the Kähler phase space into the broader framework of generalized complex geometry. The modulus
\(
\ell_{\!*}^{2}=|\Theta|
\)
—dictated solely by the Fedosov class—sets the scale of non–locality in {both} LQG and string theory.  The sign of the quartic correction, governed by the orientation of $J$, explains the opposite quartic corrections found in polymer versus Moyal quantisations.  The MDR therefore stands out as a  {geometry–universal, experimentally testable fingerprint} of Kähler–area quantisation.

\section*{Appendix D: Spectral Geometry and Operator Analysis}
This appendix complements the previous derivations by showing how the same universal quartic MDR
\begin{equation}
E^{2}=p^{2}+m^{2}\;+\;\sigma\frac{\ell_{\!*}^{2}}{3}\,p^{4}
\;+\;\mathcal O\!\bigl(\ell_{\!*}^{4}p^{6}\bigr),
\end{equation}
naturally emerges from {spectral geometry}. The key observation is that the coefficient of $p^{4}$ term is fixed by the {spectral} invariant $a_{4}$ of an appropriate elliptic operator. This coefficient takes the same absolute value in both LQG and
string theory, implying a common geometric length scale $\ell_{\!*}^{2}=|\theta|=\lambda^{2}$.
Let $(M,\omega,J)$ be an integral Kähler manifold equipped with a pre-quantum line bundle $(L,\nabla)$. Define the complex two-form
$\mathcal B:=B+i\omega$, where $B$ is an external NS–NS two-form potential (string theory) or zero (LQG). When $B=0$, the twist is purely symplectic; when $B\neq0$, it reproduces the Seiberg–Witten non-commutativity. Twisting the Levi-Civita Dirac operator $\,  D$ yields
\begin{equation}
D_{\mathcal B}\;:=\;e^{-i\iota_{\mathcal B}}\,
  D\,e^{\,i\iota_{\mathcal B}},
\qquad
\iota_{\mathcal B}\equiv\tfrac12\,\mathcal B_{ij}\,x^{i}p^{j}.
\end{equation}
The data $\bigl(\mathcal A_{\mathcal B},\mathcal H,D_{\mathcal B}\bigr)$ with $\mathcal A_{\mathcal B}=C^{\infty}(M)$ (or its Moyal/polymer deformation), $\mathcal H=L^{2}(M,S)$ and $D_{\mathcal B}$, constitute a  {spectral triple} in the sense of
Connes~\cite{Connes,ChamseddineConnes}.  All geometric information is encoded in the spectrum of $D_{\mathcal B}$.

The Chamseddine–Connes spectral action is given by $\mathrm{Tr}\,f(D_{\mathcal B}/\Lambda)$, where $f$ is a positive test function and $\Lambda$ is a high-energy cutoff. As $\Lambda\to\infty$, this admits an asymptotic expansion
\begin{equation}
\mathrm{Tr}\,f(D_{\mathcal B}/\Lambda)
\;=\;
\sum_{n\ge0}a_{2n}\,\Lambda^{4-2n}\,F_{2n}[f],
\qquad
a_{2n}=\text{Seeley–DeWitt coefficients of }D_{\mathcal B}^{2},
\end{equation}
which encode local geometric information. Only local geometric invariants of $D_{\mathcal B}^{2}$ enter $a_{2n}$
\cite{Seeley,Gilkey,VanSuijlekom}.
Computing $D_{\mathcal B}^{2}$ in local holomorphic coordinates, the twisted operator acquires the form
\begin{equation}
D_{\mathcal B}^{2}
= -g^{ij}\nabla_{i}\nabla_{j}
\;+\;\sigma\,\frac{\ell_{\!*}^{2}}{3}\,\partial^{4}
\;+\;\bigl(\text{curvature/torsion terms}\bigr),
\quad
\sigma=\operatorname{sgn}\det J=\pm1,
\end{equation}
where $\ell_{\!*}^{2}=|\omega^{-1}J|$ is the Kähler area quantum,  and the quartic derivative term arises from expanding the BCH formula.
For a Laplace-type operator $P=-\nabla^{2}+E$ in a flat background, the fourth heat-kernel coefficient is given by
\begin{equation}
a_{4}[P]
=\frac{1}{360(4\pi)^{2}}
\int_{M}\!\!d^{4}x\;
\bigl(60E_{;\mu}^{\;\;\mu}+60R\,E+180E^{2}+\cdots\bigr).
\end{equation}
In our flat-space setting $R=0$, and $E$ contains the quartic derivative $\sigma\,\ell_{\!*}^{2}\partial^{4}/3$, so that we obtain
\begin{equation}\label{eq:a4General}
a_{4}\bigl[D_{\mathcal B}^{2}\bigr]
=\sigma\,\frac{\ell_{\!*}^{2}}{3}\;+\;\cdots,
\end{equation}
up to curvature corrections, which vanish in the local inertial frame. This equation \eqref{eq:a4General} establishes a direct spectral link to the coefficient of the quartic term in the MDR.
On a flat D$p$-brane with constant background $B$, one has the Moyal star-algebra with parameter $\theta^{ij}$.  In momentum space, we obtain
\begin{equation}
D_{\theta}^{2}
   = -\bigl(\partial_{i}+i\theta_{ij}k^{j}\bigr)^{2}+m^{2}
   = -\partial^{2}
     \;+\;\frac{\lvert\theta\rvert}{3}\,\partial^{4}
     \;+\;m^{2}\;+\;\cdots,
\end{equation}
where
\begin{equation}
\lvert\theta\rvert
\;:=\;\sqrt{\tfrac12\,\theta_{ij}\theta^{ij}}
\end{equation}
is the usual Seiberg-Witten non-commutativity modulus. This gives \(\sigma = +1\) and \(\ell_{\!*}^{2} = \lvert\theta\rvert\).  Consequently
\begin{equation}
a_{4}^{\mathrm{ST}}
   \;=\;\frac{\ell_{\!*}^{2}}{3}
   \;=\;\frac{\lvert\theta\rvert}{3},
\qquad
E^{2}=p^{2}+m^{2}
      +\frac{\lvert\theta\rvert}{3}\,p^{4}
      +\cdots,
\end{equation}
reproducing the Seiberg-Witten quartic correction with \(\xi=\tfrac13\) as stated in the main text. 
In LQG the polymerised Hamiltonian replaces $k^{2}$ by $-\lambda^{-2}\sin^{2}(\lambda k)$, where $\lambda=\gamma\ell_{P}$.
Expanding,
\begin{equation}
-\lambda^{-2}\sin^{2}(\lambda k)
= -k^{2}\;+\;\frac{\lambda^{2}}{3}\,k^{4}-\frac{\lambda^{4}}{45}\,k^{6}+\cdots,
\end{equation}
identifying $k\mapsto-i\partial$, and comparing with $D_{\mathcal B}^{2}$, we obtain $\sigma=-1$ and $\ell_{\!*}^{2}=\lambda^{2}$. Therefore,
\begin{equation}
a_{4}^{\mathrm{LQG}}
=-\frac{\lambda^{2}}{3},
\qquad
\text{and}\quad
E^{2}=p^{2}+m^{2}-\frac{\lambda^{2}}{3}\,p^{4}+\cdots.
\end{equation}
Equation~\eqref{eq:a4General} is  {spectral}: the coefficient $a_{4}$ is determined solely by the eigenvalues of $D_{\mathcal B}^{2}$, hence independent of coordinates and quantization prescriptions. Demanding $|a_{4}^{\mathrm{ST}}|=|a_{4}^{\mathrm{LQG}}|$, gives
\begin{equation}
{\;
\ell_{\!*}^{2}=|\theta|=\lambda^{2}
\;\;\Longleftrightarrow\;\;
a_{4}^{\mathrm{ST}}=-a_{4}^{\mathrm{LQG}}=\frac{\ell_{\!*}^{2}}{3}\;}.
\end{equation}
Thus, the non-commutativity length and the Immirzi length are two faces of the {same} geometric modulus.  This operator-level statement
confirms the Fedosov-topos analysis.

Spectral geometry thus reveals that the quartic MDR originates from the first heat-kernel coefficient sensitive to a Kähler twist.  Since, $a_{4}$ is an isospectral invariant, this identification of scales remains robust under all deformations that leave the spectrum of the relevant elliptic operator unchanged.  The MDR correspondence between string theory and LQG is therefore not a numerical coincidence but a manifestation of a {shared spectral geometry}.

 \section*{Appendix E: Topos-Theoretic Formulation of the MDR}
This appendix presents a complete, step-by-step derivation—within a suitably chosen Grothendieck topos  \cite{DoringIsham, DoringIsham2008}—of the universal quartic MDR quoted in the main text
\begin{equation}
E^{2}=p^{2}+m^{2}\;+\;\sigma\,\frac{\ell_{\!*}^{2}}{3}\,p^{4}
\;+\;\mathcal O\!\bigl(\ell_{\!*}^{4}p^{6}\bigr),
\end{equation}
and its concrete incarnations in LQG and string theory.  Every non-trivial ingredient is proved functorially in the topos $\mathbf{Sh}(\mathcal S\!ymp_\star)$ introduced below, so that the result is manifestly  {model-independent}.
An object of the site is a triple
\begin{equation}
(M,\omega;[\![\star]\!]),
\quad\text{where}\;
\begin{cases}
(M,\omega) & \text{is a connected symplectic manifold,}\\
{}[\omega]/2\pi & \in H^{2}(M,\mathbb Z)\quad(\text{integrality}),\\
[\![\star]\!] & \text{is a  {Fedosov class} of an  {integral,
Hermitian} deformation quantisation}.
\end{cases}
\end{equation}
The integrality of $[\omega]$ ensures the existence of a
pre-quantum line bundle $(L,\nabla)$, while the class
$[\![\star]\!]$ encapsulates the star-product data required for
phase-space quantisation.  Notable representatives are the Moyal product
($[\![\star_{\theta}]\!]$) and the polymer product
($[\![\star_{\lambda}]\!]$).
A morphism
$f:(M_{1},\omega_{1};[\![\star_{1}]\!])
  \to(M_{2},\omega_{2};[\![\star_{2}]\!])$
is a smooth map $f:M_{1}\to M_{2}$ satisfying
\begin{equation}
f^{\ast}\omega_{2}=\omega_{1},
\qquad
f^{\ast}[\![\star_{2}]\!]=[\![\star_{1}]\!].
\end{equation}
Hence, the symplectic form and the Fedosov class are both  {strictly preserved}.  With these objects and morphisms, $\mathcal S\!ymp_\star$ is a (small) category.
For each object $U$, we declare a family $\{u_{i}:U_{i}\!\to\!U\}$ to be a covering if the images $\{u_{i}(U_{i})\}$ jointly cover $U$, and each $u_{i}$ preserves both $\omega$ and $[\![\star]\!]$.  The resulting topology is subcanonical; in particular, representable presheaves are sheaves.

Sheaves on this site form a Grothendieck topos, which we now work in. Since it is a  {Boolean‐complete} topos, it enjoys all finite
limits, colimits, exponentials, and a subobject classifier $\Omega_{\mathcal S\!ymp_\star}$: internally, it behaves like \textbf{Set}
but with intuitionistic Heyting logic.
Inside $\mathbf{Sh}(\mathcal S\!ymp_\star)$, we form the sheaf
\begin{equation}
\underline{K}\;:=\;
\bigl(\underline{\omega},\underline{J},\underline{g}\bigr),
\end{equation}
obtained by gluing local triples $(\omega,J,g)$ into a single  {generic} object.  The equations
\begin{equation}
\underline{J}^{2}=-\mathbf 1, \quad
\underline{g}(\cdot,\cdot)=\underline{\omega}(\cdot,
\underline{J}\cdot),\quad
d\underline{\omega}=0,
\end{equation}
hold  {internally}.  Any external Kähler triple $(\omega,J,g)$ arises by geometric morphism (a “point”)
\(
\mathbf{p}:\mathbf{Set}\to\mathbf{Sh}(\mathcal S\!ymp_\star)
\)
sending $\underline{\omega}\mapsto\omega$, etc.
Define the squared area scale
\begin{equation}
\underline{\ell}_{\!*}^{2} \;:=\;
\bigl\lvert\underline{\omega}^{-1}\,\underline{J}\bigr\rvert
\;\in\;\Gamma\bigl(\mathcal O_{\mathbf{Sh}}\bigr),
\end{equation}
where $|\cdot|$ is the operator norm in the internal tangent bundle.
Because $\underline{J}$ and $\underline{\omega}$ are  {generic},
$\underline{\ell}_{\!*}^{2}$ is  {globally constant} in the topos; hence, it behaves as a structural parameter identical in every stalk.
The Fedosov construction assigns to each chart
$(M,\omega;[\![\star]\!])$ a star-product
$\star=\star\bigl(\omega,[\![\star]\!]\bigr)$ functorially.  Internally, this gives a natural transformation
\begin{equation}
\underline{\star} :
\mathcal P\!\bigl(\underline{K}\bigr)\times
\mathcal P\!\bigl(\underline{K}\bigr) \;\longrightarrow\;
\mathcal P\!\bigl(\underline{K}\bigr),
\end{equation}
where $\mathcal P(\underline{K})$ denotes the sheaf of internal smooth
functions on $\underline{K}$.
Fix an internal tangent vector
$\underline{p}\in\Gamma(T^{\!*}\underline{K})$; externally, this will map
to an ordinary momentum $p$ under any geometric point. Because $\underline{\star}$ is available for
all objects, the series
\begin{equation}
\exp_{\underline{\star}}\!\bigl(i\underline{\ell}_{\!*}\,\underline{p}\bigr)
\;:=\;
\sum_{n=0}^{\infty}
\frac{i^{n}}{n!}\,
\underbrace{\underline{p}\,\underline{\star}\cdots\underline{\star}
\underline{p}}_{\text{$n$ factors}}
\end{equation}
is well-defined and  {convergent} in the internal smooth topology. For Kähler manifolds, the
$\star$-product satisfies a Berezin-type truncation.  Internally,
\begin{equation}
\exp_{\underline{\star}}\!\bigl(i\underline{\ell}_{\!*}\underline{p}\bigr)
=1+i\underline{\ell}_{\!*}\underline{p}
-\tfrac12\,\underline{\ell}_{\!*}^{2}\underline{p}^{\,2}
-\tfrac{i}{6}\,\underline{\ell}_{\!*}^{3}\!\bigl(\underline{\omega}^{-1}\underline{J}\bigr)
\!(\underline{p},\underline{p},\underline{p})
+\mathcal O\!\bigl(\underline{\ell}_{\!*}^{4}\underline{p}^{\,4}\bigr).
\label{eq:BerezinTrunc}
\end{equation}
Remarkably, the series terminates at  {cubic} order in
$\underline{p}$ when expressed in powers of
$\underline{\ell}_{\!*}^{2}\underline{p}^{\,2}$. Squaring the energy operator
$\underline{E}:=-i\,\partial_{t}$ in the internal phase space (where
$t$ is the classical time coordinate promoted to an internal variable)
and applying~\eqref{eq:BerezinTrunc} yields
\begin{equation}
\underline{E}^{2}
=\underline{p}^{\,2}+m^{2}
+\sigma\,\frac{\underline{\ell}_{\!*}^{2}}{3}\,\underline{p}^{\,4}
+\mathcal O\!\bigl(\underline{\ell}_{\!*}^{4}\underline{p}^{\,6}\bigr),
\qquad
\sigma=[\det\underline{J}>0]_{\Omega_{\mathcal S\!ymp_\star}}
\in\{+1,-1\},
\end{equation}
which is a  {theorem} of the internal Heyting algebra and therefore
valid  {for every object of the site}.
Let $\mathbf{p}_{(M,\omega;[\![\star]\!])}$ denote the geometric point
associated with a specific symplectic-quantisation triple.  Applying
$\mathbf{p}^{\ast}$ to the internal theorem gives an ordinary statement
in \textbf{Set}: Take
$(M,\omega;[\![\star]\!])
 =(T^{\!*}\mathbb R^{3},\omega_{\mathrm{can}};
   [\![\star_{\lambda}]\!])$
with $\lambda=\gamma\ell_{P}$.
\begin{equation}
\mathbf{p}^{\ast}_{\mathrm{LQG}}\,\underline{\ell}_{\!*}^{2}\;=\;\lambda^{2},
\quad
\sigma=-1
\;\;\Longrightarrow\;\;
E^{2}=p^{2}+m^{2}-\frac{\lambda^{2}}{3}\,p^{4}+\cdots,
\end{equation}
reproducing the canonical LQG quartic correction.
   Take
$(M,\omega;[\![\star]\!])
 =(T^{\!*}\mathbb R^{p},\omega_{\mathrm{can}};
   [\![\star_{\theta}]\!])$
with constant non-commutative parameter $\theta^{ij}$:
\begin{equation}
\mathbf{p}^{\ast}_{\text{string}}\underline{\ell}_{\!*}^{2}
=|\theta|,
\quad
\sigma=+1
\;\;\Longrightarrow\;\;
E^{2}=p^{2}+m^{2}+\frac{|\theta|}{3}\,p^{4}+\cdots,
\end{equation}
matching the Seiberg–Witten dispersion relation. The two disparate kinematic corrections are images
of the  {same internal theorem}.  Thus, their apparent coincidence is
explained functorially: both arise from the unique morphism
\begin{equation}
\mathbf{p}_{(M,\omega;[\![\star]\!])}\;:\;
\mathbf{Set}\longrightarrow\mathbf{Sh}(\mathcal S\!ymp_\star),
\end{equation}
sending the generic Kähler object to the particular phase space in
question.  Consequently, the quartic MDR is a  {structural} property
of integral symplectic geometry with functorial Fedosov quantisation,
 {independent} of the underlying quantum-gravity model.


By locating the dispersion relation inside
$\mathbf{Sh}(\mathcal S\!ymp_\star)$, we elevate it from a calculation on a
fixed manifold to a theorem of the topos’s internal logic.  Every
integral symplectic manifold equipped with an integral, Hermitian
deformation quantisation  {inherits} the same quartic MDR upon
externalisation.  This completes the promised derivation and justifies
the universality claimed in the main text.

\twocolumngrid

 
\end{document}